\documentclass[final,5p,times,twocolumn]{elsarticle}

\usepackage{amssymb}
\usepackage{amsfonts}
\usepackage{amsmath}
\usepackage{graphicx}
\usepackage{enumerate}
\usepackage[usenames,dvipsnames]{color}
\usepackage[normalem]{ulem}
\usepackage{cancel}

 \biboptions{numbers,super,sort&compress}

\journal{Journal of Crystal Growth}

\providecommand{\U}[1]{\protect\rule{.1in}{.1in}}
\newcommand{\red}{\textcolor{black}}
\def\DG{W}

\begin{document}
\begin{frontmatter}
\title{A Brownian model for crystal nucleation}
\author{Miguel A. Dur\'{a}n-Olivencia$^\star$}
\ead{maduran@lec.csic.es}
\author{F. Ot\'{a}lora}
% \ead{fermin@lec.csic.es}
\address{Instituto Andaluz de Ciencias de la Tierra. CSIC-UGR. Laboratorio de Estudios Cristalogr\'{a}ficos\fnref{lec}}
\fntext[lec]{Avda. de las Palmeras 4, 18100. Armilla, Granada, Spain}

\begin{abstract}
In this work a phenomenological stochastic differential equation is proposed for modelling the time evolution of the radius of a pre-critical molecular cluster during  nucleation (the classical order parameter).  Such a stochastic differential equation constitutes the basis for the calculation of the (nucleation) induction time under Kramers' theory of thermally
activated escape processes. Considering the nucleation stage as a Poisson rare-event,  analytical expressions for the induction time statistics are deduced for both steady and unsteady conditions, the latter assuming the semiadiabatic limit. These expressions can be used to identify the underlying mechanism of molecular cluster formation (distinguishing between homogeneous or heterogeneous nucleation from the nucleation statistics is possible) as well as to predict induction times and induction time distributions. The predictions of this model are in good agreement with experimentally measured induction times at constant temperature, unlike the values obtained from the classical equation, but agreement is not so good for induction time statistics. Stochastic simulations truncated to the maximum waiting time of the experiments confirm that this fact is due to the time constraints imposed by experiments. Correcting for this effect, the experimental and predicted curves fit remarkably well. 
Thus, the proposed model seems to be a versatile tool to predict cluster size 
distributions, nucleation rates, (nucleation) induction time and induction time statistics for a wide range of conditions (e.g. time-dependent temperature, supersaturation, pH, etc.) where classical nucleation theory is of limited applicability. 
\end{abstract}


\end{frontmatter}

\section{Introduction}
\label{sec:Introduction}

The \emph{initial (nucleation) stage} of a first-order phase transition, during which the precursors of the new phase appear as a result of random density fluctuations, still remains a challenging problem in the field of condensed matter. Throughout the last century, numerous theoretical models have been developed with the aim of describing such phenomenon. These can be grouped in two categories depending on how the energy needed to form a molecular aggregate is derived: \emph{i)} the \emph{cluster approach}, originally introduced by the pioneers of the nucleation theory \cite{book:gibbs,article:farkas,article:kaischew-stranski,article:becker-doring}; and \emph{ii)} the \emph{density-functional approach} (DFA), which was applied to nucleation first by Cahn and Hilliard.\cite{article:cahn-hilliard} 

In the cluster approach, phase transition is modelled  by the formation of unstable molecular aggregates (clusters) whose density is close to that of the new stable phase. This argument is the cornerstone of both the equilibrium and the kinetic view of nucleation; both of them consider that the aggregation of molecules requires energy, which is known as work for cluster formation $W(N)$, and such magnitude can be calculated in terms of the Gibbs Free Energy (e.g. chapter 3 of \citeauthor{book:kashchiev-1}\cite{book:kashchiev-1}). 
The mathematical expression for the work of cluster formation is very simple in the case of spherical clusters since the surface tension can be approximated by the one known between the phases involved in the process (e.g. vapor-liquid, liquid-solid) and does not depend on cluster size. Such approximation is also called \emph{capillary approximation}. In the classical nucleation theory (CNT) the work of cluster formation plays a prominent role because the equilibrium cluster size distribution is of Boltzmann type, i.e. it is exponentially proportional to the work of formation, and satisfies the law of mass action. One of the main problems of this theory is that the pre-exponential factor depends on the concentration of potential nucleation sites, an obscure parameter which cannot be determined precisely. Although CNT is easily comprehensible, in many cases it does not accurately predict the nucleation rate.\cite{article:viisanen-strey-reiss}  In the case of kinetic nucleation theory (
KNT) the pre-exponential 
factor is derived in terms of the monomer attachment and detachment rates.\cite{book:nielsen} Although the expression for the collision frequency is readily obtainable, the same is not true with the monomer desorption rate. The work of cluster formation enters in KNT via the detailed-balance condition imposed to eliminate the dependency on this unknown quantity. Considering this assumption one obtains a Fokker-Planck equation (FPE) as the time-evolution equation of the cluster size concentration, better known as the Zeldovich-Frenkel equation (e.g. chapter 9 of \citeauthor{book:kashchiev-1}\cite{book:kashchiev-1}). Hence, the equilibrium cluster size distribution can be re-written in terms of the work of formation and the monomer attachment rate yielding the kinetic counterpart of the classical equation. In this framework, the estimated values for nucleation rate are  also far from those obtained experimentally. These deviations led to the development of new theories such 
as the Density Functional Approach.

Under DFA, the state of the system is described by the number density of molecules as a function of the space coordinates, $\rho(\mathbf{r})$. No explicit boundary between the phases of higher and lower density is assumed and, hence, it cannot accurately be established whether a molecule belongs to the old or the new phase. This is also known as the diffuse interface theory of nucleation. In this regard, the work of cluster formation is now expressed in terms of the Landau potential which can be calculated by using, for example, the square-gradient approximation\cite{article:lutsko-2011b} and used as a replacement of $W(N)$ in CNT and KNT master equations.

Until recently, the above descriptions of  nucleation  have been the main tools to predict experimental results. Notwithstanding, the underlying mathematical treatment in these theories is completely deterministic, though they describe a random process. For this reason a new formulation for the nucleation theory has been developed within the statistical mechanics framework.\cite{article:ford,article:ford-review} However, this formalism cannot be used in the case of non-stationary conditions and, furthermore, it does not fit to the observed values of (nucleation) induction time.\cite{article:ford} The main motivation of this paper is to overcome both limitations, i.e. the deterministic nature of previous models and the lack of solutions for non-stationary conditions.

\subsection{Outline of the model}

In the middle of the twentieth century, Hendrik Anthony Kramers described thermally activated escape from a metastable state as \emph{Brownian motion} of a fictitious particle along a \emph{reaction coordinate}, which covers all relevant degrees of freedom that define the system state, in a (static) field of force.\cite{article:kramers} Such processes are characterized by the presence of two stable states separated by a potential barrier and can be described in two different but equivalent forms. The first is using stochastic differential equations (SDE) to characterize the time evolution of the reaction coordinate. The conjectured SDE is a Langevin-like equation\cite{article:langevin} which contains both a deterministic and a random part. The second is using the time-evolution equation of the probability density function (PDF) associated with the reaction coordinate. Indeed, this expression can be written as a Fokker-Planck equation.\cite{book:risken-1996a} Thus, as we will demonstrate later, 
using the FPE one can derive an analytical expression for the nucleation (or escape) rate in the form proposed by Arrhenius. Moreover,  
considering the nucleation process as a rare event one can obtain the PDF for the induction time.  Hence, the proposed model not only allows to reduce the problem from a $N$-dimensional to an unidimensional description but also provides analytical expressions for magnitudes such as nucleation rate under both stationary and non-stationary conditions, which demonstrates the abilities of the model. 

The aim of this work is to apply Kramers' reasoning to the problem of nucleation in order to predict the induction time statistics for both stationary and non-stationary conditions, the latter assuming the semiadiabatic limit.\cite{article:talkner-luczka,article:kim-talkner-kyun-Hanggi} We propose cluster radius, $R$, as a reaction coordinate for crystal nucleation and, hence, a stochastic differential equation is postulated for modeling time evolution of the cluster size. In this context, analytical expressions for statistics of induction time can be deduced by employing the theory of stochastic processes. This contribution allows us to better understand the randomness of experimental results and the deviation from the theoretical (classical) nucleation rate.

\section{Equation of motion for the cluster radius}

\subsection{Theoretical background}

The energy required to form (so-called work of formation of) a spherical cluster with the properties of the new phase within a supersaturated mother phase has been extensively studied.\cite{article:lutsko-2011b,article:laaksonen,article:debenedetti,article:nishioka} Although historically the usual convention has been to evaluate the work of cluster formation in terms of the increment of the Gibbs free energy, such a magnitude has to be evaluated depending on the system in terms of the corresponding thermodynamic potential. Thus, if one considers homogeneous nucleation in the context of the Grand Canonical ensemble, the work of formation will be given in terms of the Landau potential, $\Omega$. Nevertheless one can consider the problem in a closed system and, then, one has to evaluate this energy in terms of the Helmholtz free energy, $F$. In any case, Gibbs' interfacial thermodynamics can be used in order to 
derive the energetic cost of forming an spherical embryo of  the stable phase inside 
the metastable mother phase, taking into account the capillary approximation. These calculations end up in expressions that contains a volume and a surface term. 
The former consists of a free energy density term multiplied by the volume of the sphere. The latter contains the surface of the sphere multiplied by a surface tension term which generally depends on the radius. Nevertheless, as has been shown, this dependence is weak and considering the surface free energy as a constant is a very good approximation. However, computing the free energy density inside the cluster is not a straightforward calculation neither in monocomponent nor in multicomponent systems.\cite{article:laaksonen,article:debenedetti,article:nishioka} For the sake of simplicity, we will consider the approximation of such magnitude as the increment of the chemical potential\cite{book:zettlemoyer}, obtaining so the following expression for the reversible work required to form a noncritical cluster\cite{article:nishioka-1992}
\begin{equation}
 W(R)=\Delta \Omega(R)=-\frac{4\pi\,\Delta\mu}{3v_0}R^3+4\pi\sigma R^2,
 \label{eq:1}
\end{equation}
where $\Delta \mu$ is the difference between the chemical potential of the liquid and that of the solid phase, $v_0$ is the volume occupied by a molecule in the cluster of the new phase and $\sigma$ is the surface tension between the old and the new phase (considering a planar interface). At small $R$, the second term of equation (\ref{eq:1}), which is related to the formation of the surface, prevails while the first term plays a decisive role at large values of $R$. The maximum of $W(R)$ defines the critical size,
\begin{equation}
 R^\star=\frac{2v_0\sigma}{\Delta\mu}\label{eq:2},
\end{equation}
with
\begin{equation}
 W^\star=\frac{16\pi}{3}\frac{v_0^2\sigma^3}{\Delta\mu^2}\label{eq:3}.
\end{equation}

As widely known, equation (\ref{eq:1}) characterizes the equilibrium state of the system whereas the cluster radius describes the dynamic of the nucleation process. The latter magnitude is generally strongly coupled with the environmental degrees of freedom and due to such a coupling its dynamic is not deterministic but stochastic. Therefore, one can assume that $R$ plays the role of a reaction coordinate in the Kramers' theory.\cite{book:berne,article:hanggi-talkner-borkovec} Accordingly the variable cluster radius, hereafter denoted as $X_R$, can be understood as a stochastic process, i.e. $X_R(t)$ draws a random trajectory in the reduced phase space\footnote{The mathematics of such a reduction were developed in the framework of statistical mechanics\cite{book:kubo-2}} of the system (figure \ref{fig:2}). Considering these assumptions, the nucleation process is characterized by a stochastic variable which can be interpreted as the instantaneous position of a  \emph{fictitious Brownian particle}. Inspired 
by the Kramers' theory one postulates that the equation of motion for such a particle will be given by a Langevin-type equation. In fact, nucleation is nothing but a thermally activated escape process from a potential barrier. Therefore, in order to escape from the well $A=\{X_R\leq X_{R}^\star\}$, the fictitious ``random walker'' must acquire energy to overcome the energy barrier and subsequently it must again lose energy to become trapped by the attractor $B=\{X_R\geq X_R^\star\}$, i.e. a molecular cluster will fluctuate in size until escape from the well $A$ to the attractor $B$ after which the cluster will grow in a deterministic manner.

\subsection{The model}

The stochastic dynamic of the cluster radius, $X_R$, can be phenomenologically postulated by means of an \emph{overdamped} \emph{Langevin} \emph{equation},
\begin{align}
  \red{\eta_{\mathcal{E}}\,\frac{d{X}_R(t)}{dt}=-\frac{\partial W(X_R(t))}{\partial X_R}+\sqrt{2\eta_{\mathcal{E}} k_BT}\xi(t)}
 \label{eq:4}
\end{align}
where $\eta_{\mathcal{E}}$ is the friction coefficient associated with the reduced phase space (so-called \emph{reaction-coordinate viscosity}) and $\xi(t)$ denotes zero-mean, delta-correlated Gaussian white noise (GWN),
  \begin{align}
   \langle\xi(t)\rangle&=0,\nonumber\\
   \langle\xi(t)\xi(t')\rangle&=\delta(t-t')\label{eq:5}.
  \end{align}
The main reason to consider the overdamped limit (avoiding the second time derivative of $X_R$) is because it is consistent with the structure of the classical post-critical growth law. To demonstrate that we only have to consider $X_R>X_R^\star$, where the noise term is negligible in comparison to the driving force, so
$$
\frac{dX_R}{dt} \sim -\eta_{\mathcal{E}}^{-1}\frac{\partial W(X_R)}{\partial X_R}
$$
which shows the same structure as (see Eq. 2.62 of \citeauthor{book:kelton}\cite{book:kelton})
$$
\frac{d N}{dt}=-\frac{f(N)}{k_BT}\frac{\partial W(N)}{\partial N},
$$
with $N$ the number of molecules inside the cluster and $f(N)$ the attachment rate. Although inspired by the latter one may feel tented to consider a size-dependent viscosity, $\eta_{\mathcal{E}}$, we used a constant viscosity since it is in good agreement with the procedure followed in CNT to calculate nucleation rates setting $f(N)$ to $f(N^\star)$ (e.g. p.168 of \citeauthor{book:kashchiev-1}\cite{book:kashchiev-1}), with the advantage that
this simplifies the later calculations. In order to confirm that such a simplification is reasonably good, we can consider for example the case of diffusion-limited kinetics, where the monomer attachment rate is given by (see Eq. 10.18 of \citeauthor{book:kashchiev-1}\cite{book:kashchiev-1})
$$
f(N)=\gamma_N\,4\pi\,D\,\rho_\infty\,X_R(N)
$$
where $\gamma_N\simeq1$ is the sticking coefficient, $D$ is the diffusion constant and $\rho_\infty$ is the monomer number density and where the only dependence on the size enters via the radius $X_R(N)$ with $N=\frac{4\pi}{3v_0}X_R^3$. As can be observed in figure \ref{fig:R_N}, which represents  $X_R(N)$ as a function of $N$, the radius lies in the range $\sim[2,6]$ for a very wide range of $N$. Thus, one can conclude that $\eta_{\mathcal{E}}$ could be ultimately approximated as a constant to make easier the later mathematical treatment. In fact, as we will see later (Eq. \ref{eq:21-new}), if one selects $\eta^{-1}_{\mathcal{E}}=(\frac{\partial N}{\partial X_R})_{X_R^\star}^{-2}f(N^\star)/k_BT$ inspired by the previous reasoning, the nucleation rate derived from CNT is recovered except for a multiplicative factor.

It is worth to note that the friction coefficient $\eta_{\mathcal{E}}$ is an abstraction that provides information about the viscosity of the phase space, $\mathcal{E}$. Hence, there is no trivial expression for $\eta_\mathcal{E}$ in terms of the friction acting on individual molecules in the real space. Nonetheless this magnitude can be estimated, together with the mass associated with the reaction coordinate $m_X$, by using molecular dynamics.\cite{article:huang-attard} One can expect that the former will be related to the mean monomer attachment frequency for a $R$-sized spherical cluster $\overline{f(X_N)}$ since, the first term of the right hand side of equation (\ref{eq:4}) informs about the deterministic behaviour of the cluster size and the second one only can arise from the unpredictable collisions of monomers with the cluster. Indeed, $\eta_{\mathcal{E}}$ should be characterized by the thermodynamic properties of the new phase since at a fixed temperature the monomer attachment frequency depends on 
the density of the final state.

Thus the most remarkable characteristic of equation (\ref{eq:4}), comparing with the Ginzburg-Landau equation (e.g. chapter 4 of \citeauthor{book:barrat-hansen-2003-a}\cite{book:barrat-hansen-2003-a}), is the second term on the right hand side which includes not only a random variable but also the temperature of the thermal bath. Hence, this term plays the role of a fluctuating force which comprehends all degrees of freedom associated with the environment. Therefore, the equation of motion of the fictitious Brownian particle presents two contributions: \emph{i)} the deterministic force due to the free energy potential $\red{W}$, and \emph{ii)} the random force \hbox{$\widetilde{\xi}(t):=\sqrt{2\eta_{\mathcal{E}} k_BT}\xi(t)$}. 

From the stochastic differential equation (\ref{eq:4}), using the forward Kramers-Moyal expansion, it can be demonstrated that the time-evolution equation of the PDF, $\rho(X_R,t)$, is
\begin{align}
 \frac{\partial \rho}{\partial t}(X_R,t)&=\hat{\mathcal{L}}_{_{SL}}\rho(X_R,t)=-\frac{\partial j(X_R,t)}{\partial X_R}\label{eq:6},
\end{align}
where the linear differential operator $\hat{\mathcal{L}}_{_{SL}}$ is the \emph{Smoluchowski operator},
\begin{equation}
 \hat{\mathcal{L}}_{_{SL}}:=\frac{\partial}{\partial X_R}\left(\frac{1}{\eta_{\mathcal{E}}}\frac{\partial\red{W}(X_R)}{\partial X_R}+\frac{k_BT}{\eta_{\mathcal{E}}}\frac{\partial}{\partial X_R}\right)\label{eq:7},
\end{equation}
a special form of the \emph{Fokker-Planck} operator $\hat{\mathcal{L}}_{FP}$ \citep{book:risken-1996a}, and $j$ is the probability current
\begin{equation}
  j(X_R,t):=-\left(\frac{1}{\eta_{\mathcal{E}}}\frac{\partial\red{W}(X_R)}{\partial X_R}+\frac{k_BT}{\eta_{\mathcal{E}}}\frac{\partial}{\partial X_R}\right)\rho(X_R,t)\label{eq:8}.
\end{equation}
Equation (\ref{eq:6}) is also known as the \emph{Smoluchowski equation} and belongs to the family of Fokker-Planck partial differential equations. It is noteworthy that the above FPE for the proposed continuous random walker should be considered as the continuous counterpart of that derived from a discrete Brownian motion\cite{article:white-1969,book:kashchiev-1} but setting the monomer attachment rate to be the mean monomer attachment frequency, which is directly related with $\eta_{\mathcal{E}}$ by means of equation (\ref{eq:20-new}). The main advantage of having a continuous version of the PDF is that it allows us to use the tools of the continuous calculus to compute in an easy way magnitudes such as the mean first passage time without considering infinite series. Moreover, as will be immediately studied, obtaining the stationary and quasi-stationary distribution functions will be almost straightforward owing to the rules of continuous integral calculus can 
be used.

\subsection{Stationary probability density function}

The time evolution of the PDF converges to the stationary solution of equation (\ref{eq:6}) when \hbox{$t\rightarrow\infty$}. With the aid of this solution it is possible to estimate the probability to find a cluster of a given size when the time is much larger than the relaxation time $\tau_S$. This represents the time required to decay within the attractor $A$
\begin{equation}
 \tau_s\sim\left[\frac{1}{m_X}\left(\frac{\partial^2 \red{W} }{\partial^2 X_R}\right)_{X_R=0}\right]^{-1/2},
 \label{eq:9}
\end{equation}
with $m_X$ the effective mass corresponding to the fictitious particle. In such a case, i.e. when the time is larger than the time required for the spontaneous decomposition of a subcritical cluster, the non-stationary PDF converges to the stationary one
\begin{align}
\rho_{\text{st}}(X_R)&=\frac{1}{\zeta}
  \exp\left\{-\int^{X_R}\frac{1}{k_BT}\left(\frac{\partial\red{W} (X_R')}{\partial X_R'}\right)dX_R'\right\}\nonumber\\
  &=\frac{1}{\zeta}
  \exp\left\{-\frac{\red{W} (X_R)}{k_BT}\right\},\label{eq:10}
  \end{align}
where $\zeta$ is the normalization constant (see \ref{ap:1})
\begin{align}
\zeta&=\int_{0}^{\infty}\rho_{st}(y)dy\simeq
\sqrt{\frac{\pi k_BT}{2}}\frac{\exp\left\{
-\frac{\red{W}(X_R=0)}{k_BT}\right\}}{\sqrt{\left(\frac{\partial^2\red{W} }{\partial X_R^2}\right)_{X_R=0}}}\red{=\sqrt{\frac{\pi k_BT}{8\pi\sigma}}}
\label{eq:11}
\end{align}
The identity (\ref{eq:10}) is completely in accordance with the cluster size distribution of CNT\cite{article:izmailov-myerson-arnold} with $\zeta$ playing the role of the pre-exponential factor.

Nevertheless, the stationary PDF is not useful to obtain the nucleation rate since the boundary condition $X_R\geq0$ implies that the probability current must be zero at $X_R=0$, i.e. $j(0)=j(X_R)=0$, and hence at any value of $X_R$ including $X_R^\star$. Such restriction implies that the escape rate must be equal to zero due to the definition of this magnitude,
\begin{equation}
 k^+:=\frac{j}{n},\label{eq:12}
\end{equation}
with $n$ the stationary probability that the particle has not crossed the boundary $X_R^->X_R^\star$,
\begin{equation}
 n=\int_0^{X_R^-}\rho_{st}(X_R)dX_R,
  \label{eq:13}
\end{equation}
which is usually approximated by $n\sim1$. This fact shows that the stationary distribution is a good but unrealistic approximation. For this reason we shall introduce the quasi-stationary solution of the Smoluchowski equation.

\subsection{Quasi-stationary probability density function}

Now we shall assume that $\red{W/k_BT\,>\,1}$ and $T$ is constant. Under these conditions, before reaching the stationary state  (i.e. $\tau_s\ll t<\infty$)  the Smoluchowski solution remains in a quasi-stationary state and, therefore, the current probability is almost time-independent. In such a quasi-stationary state, the probability current over the top of the potential is very small near to $X_R^\star$ and the time change of the PDF is also very small. Therefore, the small value of the probability current is almost independent of $X_R$, i.e. $j(X_R,t)\sim j$. Hence, equation (\ref{eq:8}) can be written as 
\begin{align}
j&\simeq-\frac{k_BT}{\eta_{\mathcal{E}}}e^{-\red{W} (X_R)/k_BT}\frac{\partial}{\partial X_R}\left(e^{\red{W} (X_R)/k_BT}\rho(X_R,t)\right)
\label{eq:14}.
\end{align}
Considering now the boundary condition $\rho(X_R^-,t)=0$, one readily gets
\begin{align}
 \rho_{\text{q-st}}(X_R,t)&=\frac{\eta_{\mathcal{E}} j}{k_BT}e^{-\red{W} (X_R)/k_BT}\int_{X_R}^{X_R^-} dy\,e^{\red{W} (y)/k_BT}
  \label{eq:15}.
\end{align}

\section{Nucleation rates and induction time statistics}

We want to calculate the mean time that the system needs to produce a supercritical cluster, i.e. when $X_R>X_R^{\star}$. This magnitude is also called \emph{mean first-passage time} (MFPT) (or Kramers' time) and can be easily related to the induction time. In this section we present the expressions obtained for the induction time and the nucleation rate under both stationary and non-stationary conditions.

\subsection{Nucleation rate under steady conditions}

From equations (\ref{eq:12},\ref{eq:13},\ref{eq:15}) the following expression for the escape rate can be derived,
\begin{align}
 \frac{1}{k^+}&=\int_{0}^{
X_R^-} dz\,\frac{\eta_{\mathcal{E}}}{k_BT}e^{-\DG(z)/k_BT}\int_{X_R}^{X_R^-} dy\,e^{\DG(y)/k_BT}\label{eq:16}.
\end{align}
Accordingly, applying the Laplace (or Gaussian steepest-descent) method (e.g. page 124 of cite \citeauthor{book:risken-1996a}\cite{book:risken-1996a})  the following approximation can be obtained (see \ref{ap:2})
\begin{align}
k^+&=\sqrt{\frac{\partial^2\DG(0)}{\partial X_R^2}}\sqrt{\left|\frac{\partial^2\DG(X_R^\star)}{\partial X_R^2}\right|}\frac{e^{-\DG^\star/k_BT}}{\pi\eta_{\mathcal{E}} }\red{=4\pi\sigma\frac{e^{-\DG^\star/k_BT}}{\pi\eta_{\mathcal{E}} }}
\label{eq:17}.
\end{align}
The inverse of this escape rate is also known as MFPT,
\begin{equation}
 \tau_K=\frac{\pi\eta_{\mathcal{E}} }{\sqrt{\frac{\partial^2\DG(0)}{\partial X_R^2}}\sqrt{\left|\frac{\partial ^2\DG(X_R^\star)}{\partial X_R^2}\right|}}e^{\DG^\star/k_BT}.
  \label{eq:18}
\end{equation}
The nucleation rate and the induction time are given by the expressions\cite{book:barrat-hansen-2003-a}
\begin{align}
J=\frac{\rho_\infty}{\tau_K}&= \rho_\infty\,k^+=\rho_\infty\sqrt{\frac{\partial^2\DG(0)}{\partial X_R^2}}\sqrt{\left|\frac{\partial^2\DG(X_R^\star)}{\partial X_R^2}\right|}\frac{e^{-\DG^\star/k_BT}}{\pi\eta_{\mathcal{E}} }\label{eq:19-2},\\
t_{\text{ind}}&=\frac{1}{JV}=\frac{\tau_K}{\mathcal{N}_1}=\frac{\pi \eta^{\text{eff}}}{\sqrt{\frac{\partial^2\DG(0)}{\partial X_R^2}}\sqrt{\left|\frac{\partial ^2\DG(X_R^\star)}{\partial X_R^2}\right|}}e^{\DG^\star/k_BT},\label{eq:indtime}
\end{align}
where $\rho_\infty$ is the monomer number density, i.e. the equilibrium number of monomer per unit volume $\mathcal{N}_1/V$, and $\eta^{\text{eff}}=\eta_{\mathcal{E}}/\mathcal{N}_1$. As can be observed, equation (\ref{eq:19-2}) looks like the classical expression for the nucleation rate. Indeed, it can be rewritten as,
\begin{align}
J&=\rho_\infty\frac{\red{2\,k_BT}\,\left(\frac{\partial X_N}{\partial X_R}(X_R^\star)\right)}{\eta_{\mathcal{E}}}\,Z_0\,Z_D\,e^{-\DG^\star/k_BT},\label{eq:19-new}
\end{align}
with 
\begin{align}
 X_N&=\frac{4\pi}{3 v_0}X_R^3,\nonumber\\
  Z_0&=\left(\frac{1}{\red{2}\pi\red{k_BT}}\frac{\partial^2\DG(0)}{\partial X_R^2}\right)^{\frac{1}{2}}\red{=\sqrt{\frac{4\pi\sigma}{2\pi k_BT}}},\nonumber\\
 Z_D&=\left(\frac{1}{2\pi k_BT}\left|\frac{\partial^2\DG(X_N^\star)}{\partial X_N^2}\right|\right)^{\frac{1}{2}},\nonumber
\end{align}
being the number of molecules inside the $R$-sized cluster, the curvature of the energy landscape at the basin and the Zeldovich's factor (typically\cite{book:kashchiev-1}\ $10^{-2}\leq Z_D\leq1$), respectively. Therefore, the term $\red{Z_0}\frac{\red{2k_BT}}{\eta_{\mathcal{E}}}\left(\frac{\partial X_N}{\partial X_R}(X_R^\star)\right)$ sets the time scale of the phase transition. In fact,  following the classical reasoning (that the characteristic time for nucleation is determined by the attachment rate of monomers to the critical cluster) one could postulate that
\begin{equation}
Z_0\frac{\red{2k_BT}}{\eta_{\mathcal{E}}}\left(\frac{\partial X_N}{\partial X_R}(X_R^\star)\right)\sim  f(X_R^\star) \label{eq:22-22}
\end{equation}
so that
\begin{align}
  \eta_{\mathcal{E}}^{-1}&\sim Z_0^{-1}\frac{f(X_R^\star)}{\red{2k_BT}}\left(\frac{\partial X_N}{\partial X_R}(X_R^\star)\right)^{-1}\label{eq:20-new}
%  &=\widetilde{Z}_0^{-1}\frac{f(X_R^\star)}{\red{2k_BT}}\left(\frac{\partial X_N}{\partial X_R}(X_R^\star)\right)^{-2}\nonumber
\end{align}
Thus, substituting equation (\ref{eq:20-new}) into (\ref{eq:19-new}) one gets,
\begin{equation}
J\sim \rho_\infty\,Z_D\,f(X_N^\star)\,e^{-\DG^\star/k_BT}\red{=J_{\text{CNT}}}.\label{eq:21-new}
\end{equation}
Therefore, making such an interpretation of the viscosity parameter, one recovers the CNT expression of the nucleation rate\cite{book:kashchiev-1}, $J_{_\text{CNT}}$. Note that this expression of $\eta_\mathcal{E}$ is slightly different from that one we expected, i.e. $\eta^{-1}_{\mathcal{E}}= \left(\frac{\partial X_N}{\partial X_R}\right)_{X_R^\star}^{-2}f(N^\star)/k_BT$, inspired by the post-critical growth rate. In such a case, the nucleation rate would be
$$
J\sim \rho_\infty\,2\,Z_0\,Z_D\,\left(\frac{\partial X_N}{\partial X_R}\right)_{X_R^\star}^{-1}\,f(X_N^\star)\,e^{-\DG^\star/k_BT}=2\,Z_0\,\left(\frac{\partial X_N}{\partial X_R}\right)_{X_R^\star}^{-1}\,J_{\text{CNT}}
$$
which differs from the CNT expression in a multiplicative pre-exponential factor which is of the order of $Z_D$.

Nonetheless, in this work we propose a similar expression to (\ref{eq:20-new}) but considering the average value of the collision rate that an individual cluster feels, i.e. $\overline{f(X_N)}$. Although this magnitude is completely unknown we can assert that it should be lower than $f(X_N^\star)$. Therefore, our proposed expression for the nucleation rate would be as,
\begin{equation}
 J\sim  \rho_\infty\,Z_D\,\overline{f(X_N)}\,e^{-\DG^\star/k_BT},\label{eq:22-new}
\end{equation}
which should predict lower values than the classical one since, $\overline{f(X_N)}\leq f(X_N^\star)$. Nonetheless, under this assumption the nucleation rate depends on an unknown magnitude, $\overline{f(X_N)}$. This is the reason by which $\eta_{\mathcal{E}}$ must be fitted to experimental values for $t_{\text{ind}}$. Actually, one could use the fitted value of $\eta_{\mathcal{E}}$ in order to estimate the effective value of\ $\overline{f(X_N)}$\ and, then, better understand the kinetics of the phase transition. With the aid of such a value information on the underlying mechanism of cluster formation could be obtained by comparing with theoretical values obtained using the different expressions for the monomer attachment frequencies (chapter 10 of cite \citeauthor{book:kashchiev-1}\cite{book:kashchiev-1}).

\subsection{Induction time statistics under steady conditions.}

Once an estimation for the induction time has been obtained, it seems interesting to derive an analytical equation for the induction time statistics which would be helpful for a better description of the random nature of such magnitude. Thus, if nucleation is considered as a  homogeneous Poisson process\cite{article:white-1969,book:nucleation-1996,article:peters-2011,article:jiang-2011,article:limay} characterized by the escape rate, $k^+=1/\tau_K$, the induction time can be considered as a Gamma-distributed random variable with a PDF given by
\begin{equation}
 \varrho(t)=\frac{\rho_\infty}{\tau_K}\exp\left\{-\frac{\rho_\infty}{\tau_K}t\right\},
\label{eq:19}
\end{equation}
so that, 
\begin{equation}
 P(t_{\text{ind}}\leq t)=\int_0^t\varrho(s)ds=1-\exp\left\{-\frac{\rho_\infty}{\tau_K}t\right\}.
\label{eq:20}
\end{equation}
The latter equation is also called the Kramers law of the escape time statistics. According to equation (\ref{eq:18}), if we know the viscosity of the reduced phase space $\eta_{\mathcal{E}}$ then the statistics of the nucleation process can be estimated from equation (\ref{eq:20}).

Although equations (\ref{eq:18}, \ref{eq:19}, \ref{eq:20}) are good approximations, one can compare the accuracy of these expressions by numerical integration. As in the case of ordinary differential equations, there exists a huge number of techniques (stochastic integrators, SINT) for integrating a SDE as the \emph{ Euler-Maruyama method}.\cite{book:kloeden-1992-a}  In such a case, the upper limit of $X_R$ (i.e. the absorbing wall $X_R^-$) must be fixed at a value greater than $X_R^\star$ and obeying 
\begin{equation}
 |\red{W}  (X_R^-)-\red{W} (X_R^\star)|\geq2k_BT.
\end{equation}

\subsection{Time-dependent nucleation rate and induction time statistics under the semiadiabatic limit}

Let us consider the case when the potential barrier $\red{W} $ changes over time (e.g. due to changes in bulk concentration, temperature, pH, etc.\cite{article:fermin-2009}) but this change is slow compared to the relaxation time of the system $\tau_S$. According to the reasoning of  Talkner and co-workers\cite{article:talkner-luczka} the system reaches a quasi-stationary state instantaneously (semiadiabatic approximation) and therefore an analogous deduction to the developed for equation (\ref{eq:20}) can be made for unsteady conditions of $\red{W} $
\begin{equation}
 \tau_K(t)=\frac{\pi\eta_{\mathcal{E}} }{\sqrt{\frac{\partial^2\red{W} (0,t)}{\partial X_R^2}\left|\frac{\partial^2\red{W} (X_R^\star(t),t)}{\partial X_R^2}\right|}}e^{\red{W} ^\star(t)/k_BT},
  \label{eq:22}
\end{equation}
or equivalently,
\begin{equation}
 k^+(t)=\frac{\sqrt{\frac{\partial^2\red{W} (0,t)}{\partial X_R^2}\left|\frac{\partial^2\red{W} (X_R^\star(t),t)}{\partial X_R^2}\right|}}{\pi\eta_{\mathcal{E}} }e^{-\red{W} ^\star(t)/k_BT},
\label{eq:23}
\end{equation}
also called \emph{instantaneous escape rate}. Proceeding in a similar manner to the previous section, one calculates the expression for the \emph{instantaneous nucleation rate} and, hence, for the \emph{instantaneous induction time}
\begin{align}
J(t)&=\rho_\infty(t)\frac{\sqrt{2\,k_BT}\,\left(\frac{\partial X_N}{\partial X_R}(X_R^\star(t))\right)}{\eta_{\mathcal{E}}}\,Z_0(t)\,Z_D(t)\,e^{-\DG^\star(t)/k_BT},\label{eq:24-new}\\
t_{\text{ind}}(t)&=\frac{1}{J(t)V}=\frac{\tau_K(t)}{\mathcal{N}_1(t)}\label{eq:24-2-new}.
\end{align}

Equation (\ref{eq:22}) constitutes a good approximation only when the topology of the potential barrier $\red{W} (X_R,t)$ does not change, i.e. the character of its local maxima and minima must be invariant. In fact, Talkner and co-workers \cite{article:talkner-luczka} derived a more accurate expression which contains expression (\ref{eq:22}) and a second-order correction term. They called the latter as \emph{geometric correction term} because it is related to the geometric change of the barrier shape. We will not consider this second-order correction in this work.

Following a similar procedure to that of the previous section, the nucleation process can be understood now as a non-homogeneous Poisson process\cite{article:pellerey-franco-shaked-moshe-zinn-joel} and, hence, the first-passage time statistics is governed by the equation below
\begin{equation}
 \varrho(t)=\frac{\rho_\infty(t)}{\tau_K(t)}\exp\left\{-R_K(t)\right\}\label{eq:24},
\end{equation}
with
\begin{equation}
 R_K(t)=\int_0^t \frac{\rho_\infty(s)}{\tau_K(s)}ds\label{eq:25},
\end{equation}
and assuming
\begin{equation}
 R_K(\infty)=\int_0^\infty \frac{\rho_\infty(s)}{\tau_K(s)}ds=\infty\label{eq:26}.
\end{equation}
Accordingly, the distribution function of the induction time is given by
\begin{equation}
 P(t_{\text{ind}}\leq t)=\int_0^t \,\varrho(s)ds=1-e^{-R(t)}\label{eq:27}.
\end{equation}
The accuracy of our theoretical predictions is strongly related to the assumptions  made above. Consequently, the results predicted with the aid of equations (\ref{eq:22}) and (\ref{eq:27}) can be improved using a SINT. In fact, the stochastic integration is indispensable in order to obtain more realistic predictions when the hypotheses of the semiadiabatic limit are not fulfilled. Indeed, SINTs are the only tool to predict escape rates and, then, induction times under strongly unsteady conditions. As mentioned in the previous section, the instantaneous position of the absorbing wall $X_R^-(t)$ must be fixed at a value greater than $X_R^\star(t)$ and obeying the following relation
\begin{equation}
 |\red{W}  (X_R^-,t)-\red{W} (X_R^\star,t)|\geq2k_BT\label{eq:28}.
\end{equation}

\section{A qualitative analysis of the model}

At this point, a short break should be taken in order to summarize and qualitatively analyze the theoretical results obtained so far, before making use of them in the next section.

Thus far the presented model has shown the ability of reproducing the main theoretical results, such as the stationary size distribution, the nucleation rate or induction time equations, of CNT and another previous works in case of setting the viscosity parameter to be the monomer attachment rate of the critical cluster, as we discussed in section 2. Nevertheless, the major difference with previous theories is indeed that the Brownian model does not consider this as the only possibility. 
In fact, we have emphasized that there is no apparent growth mechanism to be the same as that which governs the growth of post-critical clusters. Although that could seem a disadvantage because the magnitudes mentioned above will depend on an \emph{a priori} undetermined attachment frequency, it allows us to make better predictions by deducing such a parameter experimentally under conditions when the experiments are highly reproducible and, then, using that fitted value of viscosity, $\eta$, into equations one can predict nucleation rates (or induction times) for different conditions, e.g. large values of supersaturation ratio. Therefore, following this line of reasoning, the gap between the experimental and classical predictions can be overcome, as will be shown in the next section.

Besides, another advantage of this heuristic model is to describe the randomness underlying the experiments in order to determine how reproducible they are. The induction times ineluctably have a standard deviation that cannot be reproduced by the CNT procedures. However, the Brownian model is based on a SDE and hence, the stochasticity is considered. Therefore, the present work endows the classical description with a mathematical apparatus which covers the inhere experimental deviations. That allows us to know whether or not the experimental values lie in the theoretically predicted statistics and, ultimately, knowing whether or not the experiments fulfill
our predictions.

In order to highlight and verify these claims, we will compare the results computed by using the Brownian model against those predicted by using CNT expression (i.e. Eq.(\ref{eq:21-new})). Yet more, the statistics will be tested showing a slight deviation from the experimental one, but this can be explained based on the finite number of assays.

\section{The model at work}

The Brownian model proposed in sections 2 and 3 has been tested {under fixed experimental conditions} by fitting the viscosity parameter through the expression for the induction time (\ref{eq:indtime}) to experimental results of such a magnitude measured for  hydrated calcium sulphate CaSO$_4\cdot$2H$_2$O (gypsum) in a volume of $200\,\mu L$\footnote{I. Rodr\'{i}guez-Ruiz, A.E.S. Van Driessche and J.M. Garc\'{i}a-Ruiz, data to be published.}. 
This kind of applications represents one of the main contributions of this paper because neither classical nor non-classical nucleation theories allow to follow a nucleation event in such a simple manner. As a summary, one must fit $\eta_{\mathcal{E}}$ to experimental data using equation (\ref{eq:indtime}) and apply both the SDE (Eq. (\ref{eq:4})) and equations (\ref{eq:19-2}-\ref{eq:20}, \ref{eq:24-new}-\ref{eq:27}) to predict both nucleation rates and induction time statistics under both steady and unsteady conditions.
Once we have the estimation of $\eta_{\mathcal{E}}$, one can apply equation (\ref{eq:20}) in order to calculate the induction time statistics and verify the ability to predict $P(t_{ind}\leq t)$. Moreover, with the aim of considering a finite number of assays the time evolution of the reaction coordinate $X_R$ was simulated using the Euler-Mauryama method. Furthermore, using the fitted value of $\eta_{\mathcal{E}}$ one could simulate more complicated experimental conditions where classical expressions cannot be used.

Twelve assays measuring induction time were considered at a fixed supersaturation and temperature to estimate $\eta_{\mathcal{E}}$, e.g. $\red{\overline{S}}=C/C_e=1.90$\footnote{In solution crystal growth, the supersaturation (ratio between the concetrarion $C$ and the solubility $C_e$) is used as a measure of the driving force for the phase change,
$$
\Delta \mu=k_BT\ln\left(\frac{a}{a_e}\right)\simeq k_BT\ln\left(\frac{C}{C_e}\right),
$$
where $a$ and $a_e$ are the activity and the equilibrium activity of the solute, respectively.} at $T=328K$ (the point marked with an arrow in figure \ref{fig:4}), yielding the value $\eta^{\text{eff}}\sim 2.25$. This value was computed by equaling equation (\ref{eq:indtime}) to the experimental average value at this supersaturation, i.e.
\begin{equation}
\eta^{\text{eff}}=\frac{\frac{1}{M}\sum_{k=1}^{M}t_k^{\,\text{exp}}(\overline{S})}{\frac{1}{4\sigma}\exp\left({W^\star(\overline{S})}/k_BT\right)}
\label{eq:etaeff}
\end{equation}
with $M=12$ in our case. 

\noindent{}Substituting the estimated value of $\eta^{\text{eff}}$ into equation (\ref{eq:indtime}) one can calculate a theoretical curve of predicted induction times as a function of supersaturation. The predicted curves for induction times are in very good agreement with the measured values (figure \ref{fig:4}).
To carry out the calculation of these curves, as well as the classical predictions, we need to know the surface tension. In this work 
we used\cite{book:volmer,book:kashchiev-1}
\begin{align}
\sigma&=\Psi^{1/3}(\Theta_w)\,\sigma_{_{\text{HON}}},\\
\Psi(\Theta_w)&=\frac{1}{4}(2+\cos(\Theta_w))(1-\cos(\Theta_w))^2,
\end{align}
with $\sigma_{_{\text{HON}}}=14\times10^{-3}\,$J/m$^2$ the value estimated by \citeauthor{article:alimi}\cite{article:alimi} for homogeneous nucleation (HON) in the same range of temperatures and, the contact angle $\Theta_w\simeq \pi/2$, so that $\sigma_{_{\text{HEN}}}=\sigma_{_{\text{HON}}}/2^{^{1/3}}$. This is often employed to model heterogeneous nucleation.

Moreover, as can be observed in figure \ref{fig:4}, the ratio $J/J_{_{\text{CNT}}}$ is of order $10^{-2}$  which means the effective attachment rate is $\overline{f(X_N)}\simeq 10^{-2}f(X_N^\star)$ (dividing Eq. (\ref{eq:22-new}) by (\ref{eq:21-new})). Hence, the methodology followed in CNT of approximating $f(X_N)\simeq f(X_N^\star)$  (e.g. p.168 of \citeauthor{book:kashchiev-1}\cite{book:kashchiev-1}) to obtain $J_{_\text{CNT}}$ lead to an overestimation of the nucleation rate by several orders of magnitudes.

The model also provides the necessary tools for calculating the induction time statistics by considering the Kramers law (Eqs. (\ref{eq:20}) and (\ref{eq:27})). The same set of twelve experimental values for induction time were used in order to calculate the experimental cumulative distribution function (blue triangles in figure \ref{fig:5}). Using the estimated value of $\eta^{\text{eff}}$ into equation (\ref{eq:20}), $P(t_{\text{ind}}\leq t)$ (solid orange line in figure \ref{fig:5}) was obtained. As can be observed, the theoretical prediction does not fit as expected to the experimental statistics. Such a disagreement between the predicted and the observed curves is not due to the assumptions made to integrate the escape rate but to the fact that in experiments we truncated the statistics to the longest time observed, $t_{\text{max}}^{\text{exp}}$, i.e. no nucleation events were recorded after an arbitrary time $t_{\text{max}}^{\text{exp}}$ corresponding to the duration of the experiment.
 So the experimental $t_{\text{ind}}$ values were biassed towards small values. Using a SINT (Euler-Maruyama) to simulate with equation (\ref{eq:4}) a finite number of assays with an upper limit equal to the experimental observation time (i.e. if a simulation exceeds the upper limit $t_{\text{max}}$, then such assay is not considered and another one starts) the simulated results (green squares in figure \ref{fig:5}) are much closer to the experimental curve.

\section{Conclusions}
In this work a stochastic differential equation was presented as the equation of motion of the classical order parameter in the classical nucleation theory, i.e. the cluster radius. This SDE for modelling the time evolution of the radius of an individual cluster was used to obtain theoretical equations to predict (nucleation) induction times and its statistics. These equations are applicable to the often used crystallization setups in which supersaturation changes over time and can be used to identify the underlying mechanism of cluster formation by fitting the measured nucleation rates to equations (\ref{eq:19-new})-(\ref{eq:20-new}) and (\ref{eq:22-new}), as well as to predict induction times (Eqs. (\ref{eq:indtime}) and (\ref{eq:24-2-new})) and induction time distributions (Eqs. (\ref{eq:19})-(\ref{eq:20}) and (\ref{eq:24})-(\ref{eq:27})). 
We present here not only an application of the model but also a method to obtain 
theoretical 
and simulated predictions of both induction times (and hence of nucleation rates) and cumulative distribution functions at different concentrations and temperatures.
The first test of the stochastic model against experimental data reveals its potential ability for calculating nucleation rates and induction time statistics. Both the analytical and the numerical results predicted by the Brownian model seems to be in good agreement with the experimental data. Hence, the interpretation of the nucleation stage as an escape process could be an optimal tool to study deeper 
problems of the first-order phase transitions. However the theoretical approximations usually offers worse results than the simulations of the SDE by using  stochastic integrators due to the experimental truncation of the statistics to the longest time observed. Additional work is in progress to deduce analytical expressions for nucleation rate, induction times and induction time distributions when neither stationary nor semiadiabatic limit can be considered.

\section*{Acknowledgements}

This research was supported by Ministerio de Ciencia e Innovaci\'{o}n,
FPI grant BES-2010-038422 (project AYA2009-10655). The authors wish to thank I. Rodr\'{i}guez-Ruiz for providing us unpublished results. We are also immensely grateful to Prof. J.F. Lutsko, Dr. J.M. Delgado-L\'{o}pez, Dr. A.E.S Van Driessche and Dr. J.A. Gavira for their comments on an earlier version of the manuscript.

\appendix
\section*{Appendix}
\section{Approximation of the normalization constant}
\label{ap:1}

Let us consider that $\red{W} ^\star/k_BT$ is large and that $k_BT$ is very small. Therefore $\rho_{st}$ becomes very small for values of $X_R$ appreciably different from $X_R=0$. In this case $\red{W} $ can be expanded according to Taylor's theorem as
\begin{align}
  \red{W} (X_R)&=\sum_{k=0}^{\infty}\frac{1}{k!}\left(\frac{\partial^{\,k}\red{W} }{\partial X_R^k}(0)\right)X_R^{\,k}\\
&\approx \red{W} (0)+\frac{1}{2!}\left(\frac{\partial ^2\red{W} (0)}{\partial X_R^2}\right)X_R^2\nonumber,
\end{align}
and thus one gets the following approximation
\begin{align}
 \zeta&\approx\int_0^\infty \exp \left\{ -\frac{1}{k_BT}\left[ 
      \red{W}(0) +\frac{1}{2}\red{W} ''(0)X^2
	\right] \right\}dX\nonumber\\
  &=e^{-\red{W} (0)/k_BT}\int_0^\infty\exp \left\{ -\frac{1}{2k_BT}\red{W} ''(0)X^2
	\right\}dX\nonumber\\
  &=e^{-\red{W} (0)/k_BT}\left.\frac{\sqrt{\frac{\pi k_BT}{2}}\,\text{erf}\left(\frac{\sqrt{\red{W} ''(0)}\,X}{\sqrt{2k_BT}}\right)}{\sqrt{\red{W} ''(0)}}\right|_{X=0}^{X=\infty}\nonumber\\
  &=\sqrt{\frac{\pi k_B T}{2}}\frac{e^{-\red{W} (0)/k_BT}}{\sqrt{\red{W} ''(0)}},
\end{align}
with
$$
\red{W} ''(X)\equiv\frac{\partial^2\red{W} (X)}{\partial X^2}
$$
denoting the second derivative with respect the reaction coordinate $X$.

\section{Integration of the escape rate equation}
\label{ap:2}
Whereas the main contribution to the first integral in equation (\ref{eq:16}) stems from the region around $X_R=0$, i.e. close to the minimum of the barrier, the main contribution to the second integral stems from the region around $X_R^\star$. Therefore, considering  Taylor's expansion of $\DG$ around its minimum and maximum,
\begin{align}
\red{W} (X_R)&\approx \red{W} (0)+\frac{1}{2!}\left(\frac{\partial^2\red{W} (0)}{\partial X_R^2}\right)\times\nonumber\\
  &\times(X_R-0)^2+\mathcal{O}(X_R^3)\label{eq:B1},\\
\red{W} (X_R)&\approx \red{W} (X^\star)-\frac{1}{2!}\left|\frac{\partial^2\red{W} (X_R^\star)}{\partial X_R^2}\right|\times\nonumber\\
  &\times(X_R-X_R^\star)^2+\mathcal{O}((X_R-X_R^\star)^3)\label{eq:B2}.
\end{align}
and substituting equations (\ref{eq:B1}) and (\ref{eq:B2}) into equation (\ref{eq:16}), one obtains the escape rate equation,
\begin{equation}
\frac{1}{k^+} \simeq \frac{\pi\eta_{\mathcal{E}}}{2}e^{\DG^\star/k_BT}\frac{1}{\omega_0\omega^\star}\left\{\text{erf}\left[\frac{\omega^\star(X_R-X_R^\star)}{\sqrt{2k_BT}}\right]\right\}_{X_R=0}^{X_R^-},
\end{equation}
with
\begin{align}
 \omega_0&=\sqrt{\frac{\partial ^2\DG(0)}{\partial X_R^2}}\red{=\sqrt{4\pi\sigma}},\nonumber\\
 \omega^\star&=\sqrt{-\frac{\partial^2\DG(X_R^\star)}{\partial X_R^2}}\red{=\sqrt{4\pi\sigma}}.\nonumber
\end{align}
In the case that $X_R^\star>1$ and $(X_R^--X_R^\star)>1$ one can consider
$$
\left\{\text{erf}\left[\frac{\omega^\star(X_R-X_R^\star)}{\sqrt{2k_BT}}\right]\right\}_{X_R=0}^{X_R^-}\simeq2
$$
as a good approximation. Finally, one readily obtains the desired expression for the escape rate
\begin{equation}
k^+\simeq \frac{\omega_0\omega^\star}{\pi\eta_{\mathcal{E}}}e^{-\DG^\star/k_BT}\label{eq:B4}.
\end{equation}
Nevertheless, the accuracy of the result given by equation (\ref{eq:B4}) depends on the goodness of the approximations assumed above.

%%\bibliographystyle{model1-num-names}
%%\bibliography{bib-red}

\renewcommand{\thefigure}{\arabic{figure}}

\begin{figure*}
\begin{center}
\includegraphics[width=0.30\textwidth]{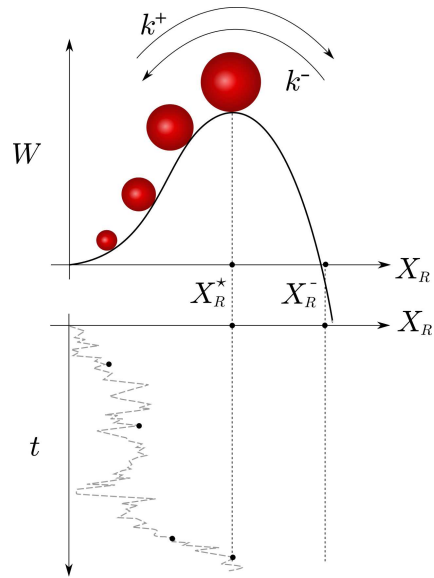}
\end{center}
\caption{Free energy barrier, $\red{W}$, as a function of the cluster radius, with a metastable state at $X_R=0$ . Escape occurs via the forward rate $k^+$. Red balls represent spherical clusters of the new phase with radius $R$ and black spots represent different values of the reaction coordinate $X_R$ which follows a Brownian motion immersed in an external field of force derived from $\red{W}$. The absorbing wall has been denoted as $X_R^-$.\label{fig:2}}
\end{figure*}

\begin{figure*}
  \begin{center}
    \includegraphics[width=0.4\textwidth]{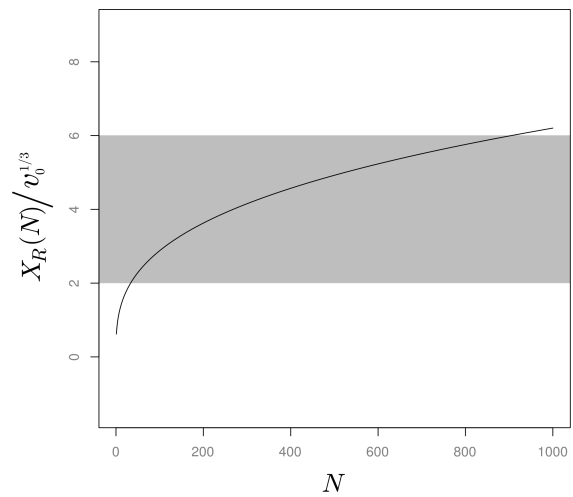}
  \end{center}
  \caption{Cluster radius represented as a function of the number of molecules. The shadow area emphasizes the fact that this magnitude can be treated up to first order of approximation as a constant in a very wide range of values for $N$.\label{fig:R_N}}
  \end{figure*}

\begin{figure*}
\begin{center}
 \includegraphics[width=0.43\textwidth]{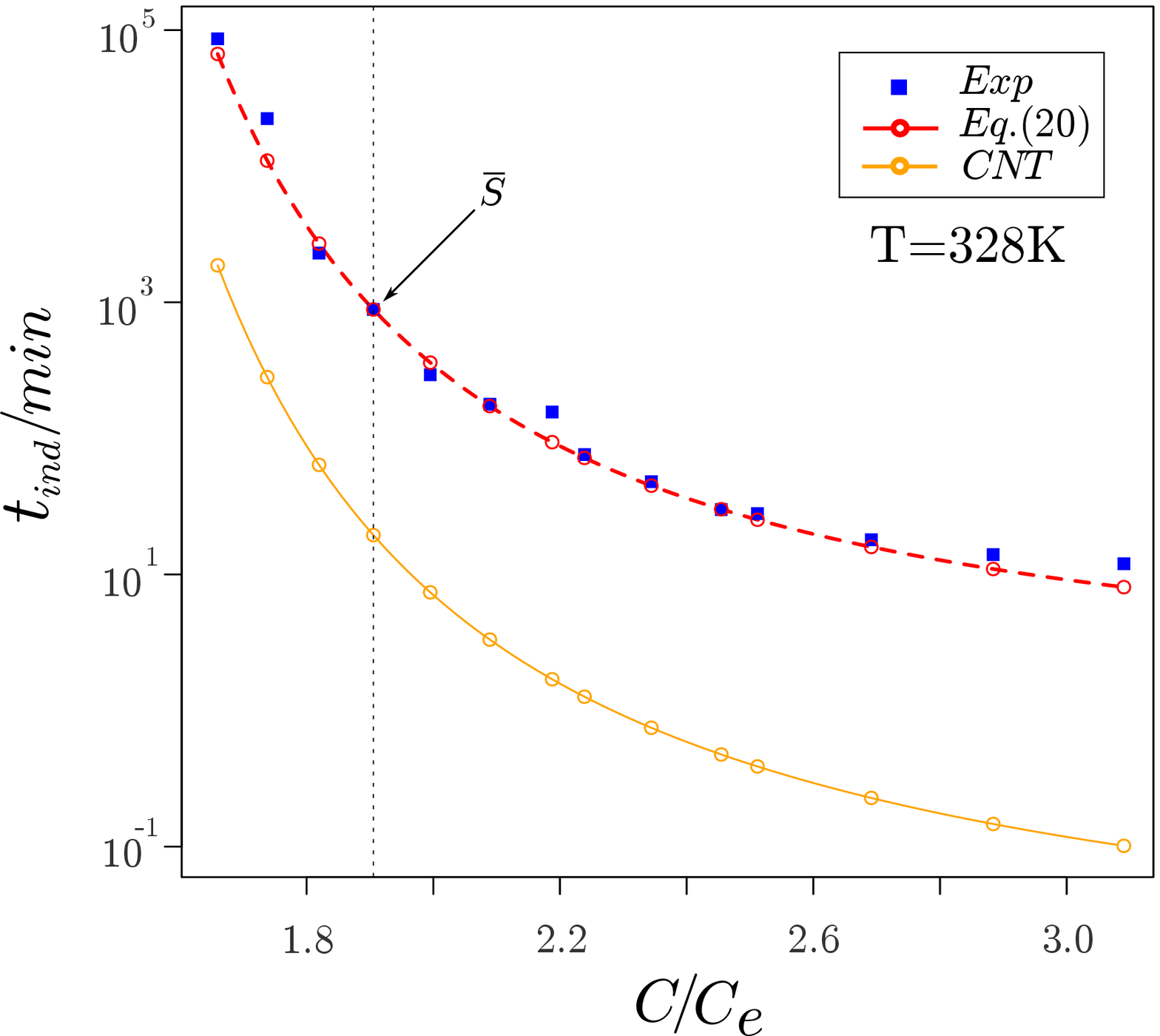}\hspace{0.5cm}\\
 \vspace{0.5cm}
 \includegraphics[width=0.43\textwidth]{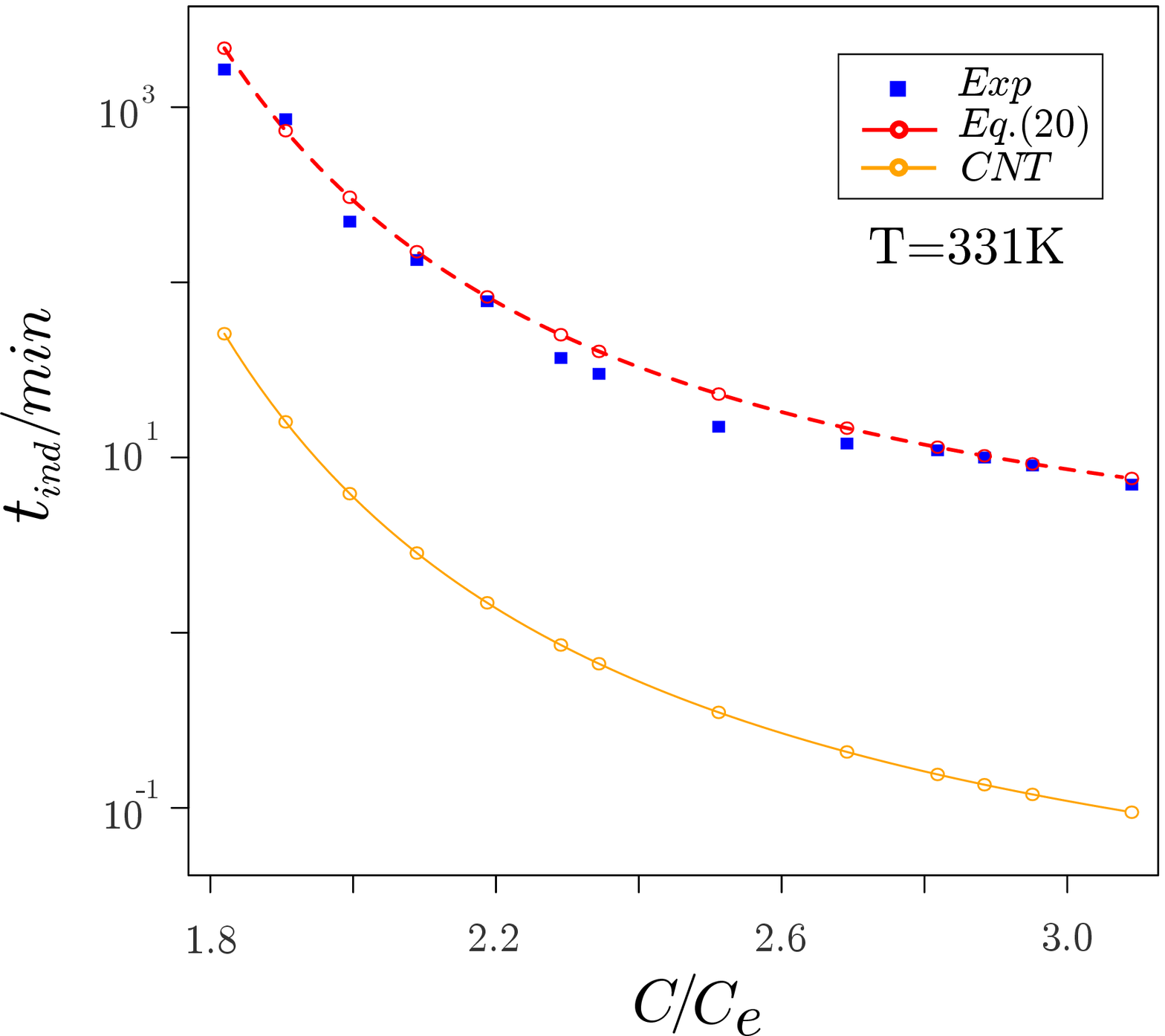}
\end{center}
  \caption{Experimental and theoretical values of $t_{ind}$ as a function of supersaturation $C/C_e$ at $T=328\,K$ (left panel) and  $T=331\,K$ (right panel). Dashed (red) lines correspond to induction times predicted by using equation (\ref{eq:18}) and the estimated friction coefficient $\eta^{\text{eff}}\sim2.25$. Each experimental value represent an average of twelve assays. Orange (solid) lines are the estimations computed by using CNT, i.e. equation (\ref{eq:21-new}), assuming the case of diffusion-limited kinetics in $f(X_N^\star)$.}\label{fig:4}
\end{figure*}

\begin{figure*}[t]
  \begin{center}
 \includegraphics[width=0.45\textwidth]{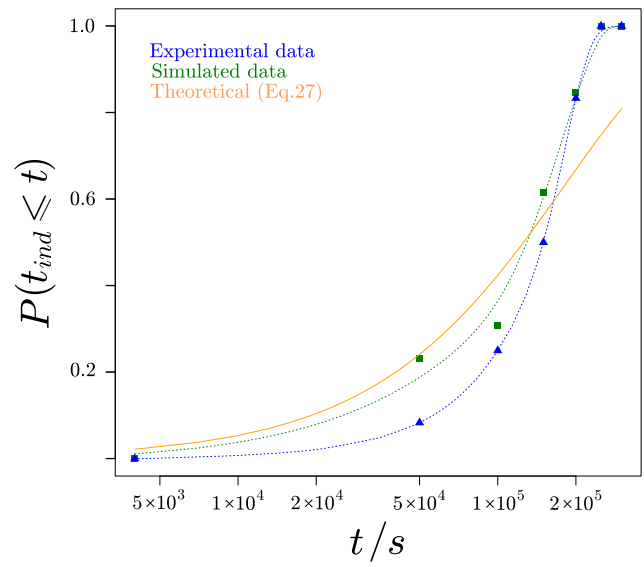}
   \end{center}
 \caption{The cumulative distribution $P(t_{ind}\leq t)$ as a function of time: \emph{i)} experimental data (blue triangles) and \emph{ii)} simulated results (green squares). Solid orange line represents the  cumulative distribution function given by equation (\ref{eq:20}).}\label{fig:5}
\end{figure*}

\end{document}